\documentclass[%
 reprint,
 amsmath,amssymb,
 aps,
]{revtex4-1}
\usepackage{comment}
\usepackage{xcolor}

\usepackage{graphicx}
\usepackage{dcolumn}
\usepackage{bm}
\usepackage{braket}
\preprint{APS/123-QED}
\begin{document}

\title{Cavitation instabilities in amorphous solids via secondary mechanical perturbations}
\author{Umang A. Dattani$^{1,2,a}$, Rishabh Sharma$^{3,a}$,  Smarajit Karmakar$^{3,*}$ and Pinaki Chaudhuri$^{1,2}$}
\email{smarajit@tifrh.res.in} 
\email{pinakic@imsc.res.in}
\affiliation{$^1$The Institute of Mathematical Sciences, CIT Campus, Taramani, Chennai 600113, India\\
	$^{2}$Homi Bhabha National Institute, Anushaktinagar, Mumbai 400094, India\\
	$^{3}$Tata Institute of Fundamental Research, 
	36/P, Gopanpally Village, Serilingampally Mandal,Ranga Reddy District, 
	Hyderabad, 500046, Telangana, India}
\thanks{$^a$ equally contributed} 






\begin{abstract}
Amorphous solids are known to fail catastrophically and in some situations, nano-scaled cavities are believed to play a significant role in the failure. In a recent work, using numerical simulations, we have shown the correspondence between cavitation under uniform expansion of amorphous solids and the yielding under shear. In this study, we probe the stability of spatially-homogeneous states sampled from expansion trajectories to alternate modes of driving, viz. macroscopic cyclic shear or local random deformation via activity. We find that, under cyclic shear and activity, the cavitation instabilities can occur in expanded states at much higher densities than under pure uniform-expansion, and the shift in density
is determined by the magnitude of the secondary deformation. We also show that barriers to cavitation on the energy landscape are much smaller for cyclic-shear and activity than seen under expansion. Further, we also analyse the spatial manifestation of cavitation and investigate whether large scale irreversible plasticity can set in due to the combination of expansion and the secondary deformation. Overall, our study reveals the interplay between expansion and other deformation modes leading to cavitation instabilities and the existence of abundant relaxation pathways for such processes.          
\end{abstract}
\maketitle
\section{Introduction}

The phenomenon of cavitation corresponds to the sudden formation as well as expansion of a void or bubble in a material that is experiencing a 
mechanical tension \cite{franc2006fundamentals, young1999cavitation}. Theoretically, cavitation is perceived to occur when the glass-gas coexistence regime of the material's phase diagram is 
accessed during mechanical loading. In liquids, cavitation has been extensively studied  \cite{rayleigh1917viii, plesset1977bubble, maris2000negative, caupin2006cavitation}, especially since collapsing bubbles often inflicts damages to
the adjoining surfaces. On the other hand, in solids, while cavitation has been observed quite long back \cite{gent1959internal, gent1970surface}, 
a proper physical understanding of the  processes at play remains missing. In recent times, there is an increased interest in the phenomenon given its significance 
in the context of failure of amorphous solids via crack propagation \cite{bouchaud2008fracture, murali2011atomic, singh2016cavitation}. Recent experiments have clearly demonstrated that emergence of isolated cavities and subsequent merger of 
these micro-cavities eventually leads to complete failure via fracture \cite{CavitationSciadv.abf7293}. Further, cavitation is often cited as the source of 
damages in soft disordered solids like biological tissues \cite{raayai2019intimate,barney2020cavitation}. Also, cavitation rheology is now an emerging protocol to characterize stability of soft materials to localized perturbations \cite{zimberlin2007cavitation}. 
Therefore, to design better materials which can sustain against such failure, we need to have a proper physical understanding of the microscopic
processes leading to cavitation in solids.  

Given the wide range of applications in which amorphous solids are used, starting from the nanoscale devices to large appliances, understanding 
the failure processes of amorphous solids has been an area of active research over the last few decades \cite{schuh2007mechanical, Rodney2011, bonn2017yield, RevModPhys.90.045006}. Typically, hard amorphous solids 
(e.g. metallic glasses, silicates, polymer glass etc.) are known to fail catastrophically, i.e. an usual brittle response which leads to fracture. Till now, 
much of the focus has been on shear response of dense glasses, where significant progress has been made to characterize the detailed elasto-plastic 
response of the material including development of mesoscale elasto-plastic models to understand yielding under shear on a mesoscopic length scale 
\cite{RevModPhys.90.045006,SrikanthPRX,ContinuumModelling}. Not much has been explored in the context of  cavitation in such materials under 
uniform expansion or uniaxial deformation process. There are a handful of numerical studies that recently focused on understanding the microscopic 
mechanisms of how cavitation occurs during volumetric strain \cite{FalkCavitation,PinakiCavitation,Pablo1,KallolCavitation}, wherein the emergence of
the cavities has been demonstrated as a yielding process \cite{PreviousWork}. In fact, it has been shown \cite{PreviousWork} that cavitation instability under
uniform volumetric expansion in a athermal quasistatic limit is a saddle-node bifurcation in the potential energy landscape where a stable 
minimum becomes unstable by merging with the nearest saddle during the expansion process. Such a process manifests itself via vanishing of an 
eigenvalue of an appropriate dynamical matrix (Hessian matrix) of the energy function. This results suggests an unified picture of failure of amorphous 
solids on the energy landscape under various deformation protocols and indicates a clear possibility of cross coupling between these deformation 
modes during a natural deformation process in a material which certainly involves nearly all deformation paths like shear, volumetric expansion, uniaxial 
tension etc. Ramification of these cross coupling can be detrimental to the stability of the materials under mechanical loading. In fact, recent studies have probed 
the interplay of these modes in fracture propagation from a  seeded void \cite{LernerBouchbinder}. Thus, the fundamental question of how the cavitation 
instabilities spontaneously occur in a homogeneous solid via various combinations of loading needs a systematic study. Further, whether the location of 
the instability and thereby the phase diagram gets altered during these diverse non-equilibrium processes needs to be explored.  


In this work, using numerical simulations, we probe cavitation in states generated during the uniform expansion of a dense amorphous solid, when 
subjected to secondary modes of mechanical deformation.  We use two different secondary modes: (i) athermal quasistatic cyclic shear which is a 
macro-scale perturbation, and (ii) random local perturbation via self-propelled active particles, which is a perturbation at micro-scale. In the first scenario 
it was done under athermal quasi-static conditions whereas in the second case, we kept the solid a temperature well below the calorimetric glass transition
temperature ($T_G$) defined as a temperature where the relaxation time in our simulation exceeds $10^6$ in the appropriate reduced units. We 
demonstrate that irrespective of the form of secondary perturbation, cavitation happens at densities higher than where it occurs for a pure expansion 
protocol. Thus, this indicates that cavitation instabilities couple to bulk shear and local forcing at these higher densities in conjunction with the tension 
generated via expansion. We summarize our findings in the form of a phase diagram (see Fig.\ref{fig1}), wherein the magnitude of the secondary forcing 
and the density are the control parameters. Further, we find that cavitation by itself may not lead to large-scale plasticity and flow. Rather, the cavitated 
states that emerge via the secondary forcing can be in an arrested or aging glassy states, depending on whether they are in an athermal or thermal conditions, 
respectively. Only if the magnitude of the forcing is large enough, an irreversible plasticity occurs via avalanche-like spatial structures, as observed in 
the case of athermal cyclic shear. Overall, our study reveals that the addition of the secondary mechanical perturbations provide pathways for accessing 
lurking cavitation instabilities when the amorphous solid is under tension due to an ongoing expansion process. 

The rest of the paper is organized as follows. We first discuss the deformation protocol we used to study the coupling between different deformation 
processes with complete details of the simulation methodology is described in the Methods section. We then discuss the results and conclude the paper
with implication of these  results on understanding microscopic failure mechanisms of amorphous solids under deformation with the aim to 
develop fundamental knowledge for preventing catastrophic failure in these materials and designing better materials for future applications.  


\section{Models and Methods}

\subsection{Model and Simulation Details}
For our study, we use the well-characterized two-dimensional model binary Lennard-Jones mixture (KABLJ) which has $65:35$ composition of the two species, labelled $A$ and $B$, with interaction parameters $\sigma_{AA}=1.0$, $\sigma_{BB}=0.88$, $\sigma_{AB}=0.8$, $\epsilon_{AA}=1.0$, $\epsilon_{BB}=0.5$, $\epsilon_{AB}=1.5$ \cite{KobAndersen2D}. We smoothen the Lennard-Jones potential up to first two derivatives. The interactions between particles $i$ and $j$ take the following form:
\begin{equation}
V_{mn}(r_{ij})=4\epsilon_{mn}\left[\left( \frac{\sigma_{mn}}{r_{ij}} \right)^{12} - \left(\frac{\sigma_{mn}}{r_{ij}} \right)^6\right] + u(r_{ij})
\end{equation}
where,
\begin{equation}
u(r_{ij})=C_{0} + C_{2}\left(\frac{r_{ij}}{\sigma_{mn}}\right)^2 + C_{4}\left(\frac{r_{ij}}{\sigma_{mn}}\right)^4
\end{equation}
Here, $m$ and $n$ would correspond to either of the labels $A$ or $B$ which signify the type of particles.
The constants $C_0$, $C_2$ and $C_4$ are determined by requiring the potential and its first two derivates to be zero at the cutoff $r_c=2.5\sigma_{mn}$. The simulations reported in the paper are performed on a system size of $N=5000$ with periodic boundary conditions.\\

\noindent{\bf Preparation of Initial states: }
Cyclic shear: To prepare initial states for our study, we first equilibriate the system at $T=1.0$ (in LJ units), which is in 
the high temperature liquid regime, followed by cooling at a constant rate of $\dot{T}\approx 10^{-4}$ per MD timestep to $T=0.01$, followed by a quech to $T=0$. For the active dynamics, the initial states are prepared by cooling the high temperature liquid at $T=1.0$ to $T=0.01$ at a cooling rate $\dot{T}=2.5\times 10^{-7}$ per MD timestep. To check the dependence of the phenomena on preparation-protocol, the results have been cross-checked for slower cooling rates that generate more stable (lower-energy) amorphous solids too. 

\subsection{Protocols for mechanical deformations}

Here, we outline the various protocols of mechanical deformation that we have used. Expansion is the primary deformation mode, starting from
the high density amorphous solid. From this expansion trajectory, be it under athermal conditions or at finite temperatures, initial states are generated at different densities upon which the secondary deformation is imposed. The details of the primary and secondary deformations are outlined below.\\

\noindent{\bf Athermal quasistatic expansion (AQE): }
The solid at $T=0.01$ is quenched to $T=0$ by conjugate gradient energy minimization. The solid is then subjected to athemal quasistatic expansion (AQE) \cite{Pablo1, Pablo2, PreviousWork} where, in each step, a constant volume strain is applied on the system by rescaling the length of the box by a factor $(1+\epsilon)$ along with affine transformation of particle coordinates, followed by minimization of the energy of this strained configuration using the conjugate gradient algorithm. We used $\epsilon=5\times 10^{-5}$. 
The configurations from the AQE trajectory, at different densities, are used as initial states for the athermal quasistatic cyclic shear procedure detailed below. These simulations are done using the open source software LAMMPS \cite{LAMMPS}.\\

\noindent{\bf Expansion at finite-temperature:}
The solid cooled down to $T=0.01$ is subjected to a uniform volumetric strain by scaling the length of the box as well as the particle coordinates by a factor of 1 + $\epsilon$ in each time-step ($dt=0.005$), using $\epsilon= 10^{-8}$. 
Thermostating is achieved by using Nos\'e-Hoover chains. The configurations sampled across this finite-temperature expansion trajectory are used as initial states for imposing the drive at the microscopic scale via activity as detailed below.\\


\noindent{\bf Athermal Quasistatic cyclic shear (AQCS): }
In recent times, the response of amorphous solids to such a shear protocol has been studied in details \cite{AQCS, ParmarSastryNature}. In this protocol, in each step, we apply a shear strain of magnitude $\delta \gamma\approx10^{-4}$ by deforming the simulation box and remapping the particle coordinates, using LAMMPS. The system is then allowed to relax via energy minimization using conjugate gradient(CG) algorithm. One cycle of AQCS consists of the following steps: (a) Driving of the system to a shear strain $\gamma=\gamma_{max}$, using quasistatic steps of magnitude $\delta \gamma$; (b) Reversal of direction of applied shear strain upto a strain of $\gamma=-\gamma_{max}$ (c); Another reversal in direction of the applied strain until the strain reaches $\gamma=0$. Several such cycles are conducted and the response of the system is studied.\\

\noindent{\bf Active dynamics: }
To impose driving via activity, we implement run and tumble dynamics. To a small fraction (2\%) of the particles, chosen at random, a random active force $\vec{f_0} = f_0(k_x\hat{x} + k_y\hat{y})$ is imposed.   By restricting the value of k's to $\pm 1$, the forces added act only along any of the four diagonal directions. This 4-state clock model is the 2 dimensional analog to the 8-state model used in \cite{mandal2016}. The directions of active forces are randomised after a persistence time $\tau_p=1$. In addition, the conditions $\sum k_x =0$ and $\sum k_y =0$ are imposed which in-turn ensure that the net momentum added to the system due to the active forces was zero at all times.\\

\subsection{Measurements} 

Here, we outline how different observables have been measured, in our analysis.\\

\noindent{\bf Coarse-graining density field:}
For our analysis, we construct a coarse-grained density field by using the method described in Ref.\cite{TestardBerthierKob} via a coarse-graining radius of $r=1.5\sigma$. In this coarse-grained density field, a region of density $\rho<0.5$ is considered as a cavity for marking the points in the phase diagram.\\

\noindent{\bf Overlap function:} 
To check for the glassiness of the spatially inhomogeneous states that emerge after cavitation has happened, we compute the overlap correlation function with respect to a state after the cavity is formed. The correlation function is defined as
\begin{equation}
q(t) = \frac{1}{N} \sum_{i=1}^N \theta(a - |r_i(t) - r_i(t_0)|) 
\end{equation}
Here $\theta(x)$ is the unit step function, and we choose a = 0.3 to allow for small thermal vibrations about mean. Averaging
is done over an ensemble of trajectories orginating from independent initial states.
We show data at $\rho = 1.108$ where we sample initial states from the expansion trajectory  and impose active dynamics, 
The overlaps were then measured with respect to the configurations at time $t_0$, corresponding to $5\times10^6$ MD steps after the imposition of active dynamics, by when the cavity has emerged for all the trajectories within the ensemble.  \\

\noindent{\bf Mobility maps:}
To get spatial information about the extent of displacemets at the local scale, due to the imposed secondary deformation, we construct mobility maps. For each particle within the system, we compute the squared distance travelled at time $t$ (or cycle number) relative to the initial state ($t=0$), i.e the time at which secondary deformation is imposed on a state sampled from the expansion trajectory,  
\begin{equation}
\Delta^2_{i}(t)= \left[ \bm{r}_i (t) - \bm{r}_i (0) \right]^2
\end{equation}
We then colour each particle in the initial configuration ($t=0$) as per the magnitude of $\Delta^2_{i}(t)$.

\section{Perturbation protocol}

\begin{figure}
	\centering
	\centerline{\includegraphics[width=1.2\linewidth]{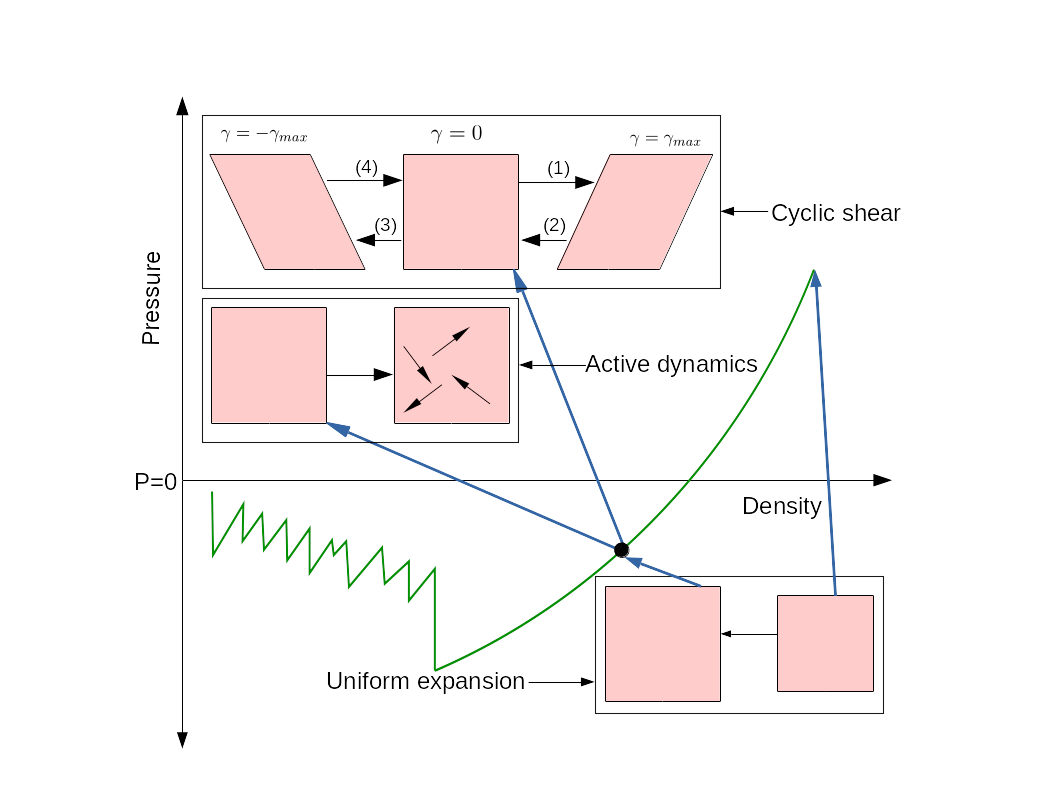}}
	\caption{A schematic diagram of the protocol followed in our numerical experiment: states generated during the expansion from a dense spatially homogeneous
		amorphous solid are subjected to secondary deformations, either in the form of a cyclic shear or random active forcing.}
	\label{fig0}
\end{figure}

In Fig.\ref{fig0}, we illustrate a schematic of the protocol that we follow, in our study. When an amorphous solid having attractive interactions is expanded, 
either under athermal or thermal conditions, it's pressure decreases and becomes negative, and then cavitates via a pressure jump; this is shown via 
the  pressure ($P$) vs density ($\rho$) line (shown in blue) in the schematic diagram. The secondary deformation protocol is then imposed on states 
generated along that $P$ vs $\rho$ trajectory. In this study, we use two different types of secondary deformations. In one case, we apply a macroscopic perturbation in the form of a quasistatic 
cyclic shear \cite{AQCS} having a certain strain amplitude ($\gamma_{max}$) which becomes a control parameter of our study.  In the second case, we apply random 
active forcing on a subset of the particles, with net applied force being zero; this is labeled as active dynamics; the magnitude
of the active forcing ($f_0$) is the control parameter in this case. Earlier studies have demonstrated how the random active forcing introduces local shear within the system  \cite{Manning, mandal2020extreme}.
The details for both the methods is discussed in the Methods section. Note that, recent  studies have explored the response of amorphous solids to athermal quasistic cyclic shear \cite{ParmarSastryNature,SrikanthPRX} as well as to random active forcing \cite{Manning, mandal2020extreme}, at high density, and demonstrated that the material yields, i.e. undergoes largescale irreversible plasticity, when the forcing is large enough. Here, we are using these protocols as secondary perturbations in conjunction with the uniform expansion, over a large density range where the pressure experienced by the 
material varies from positive to negative, to study the mechanical response of the amorphous solid, with specific focus on the phenomenon of cavitation.

\section{Results}
\begin{figure*}
	\centering
	\includegraphics[width=0.9\linewidth]{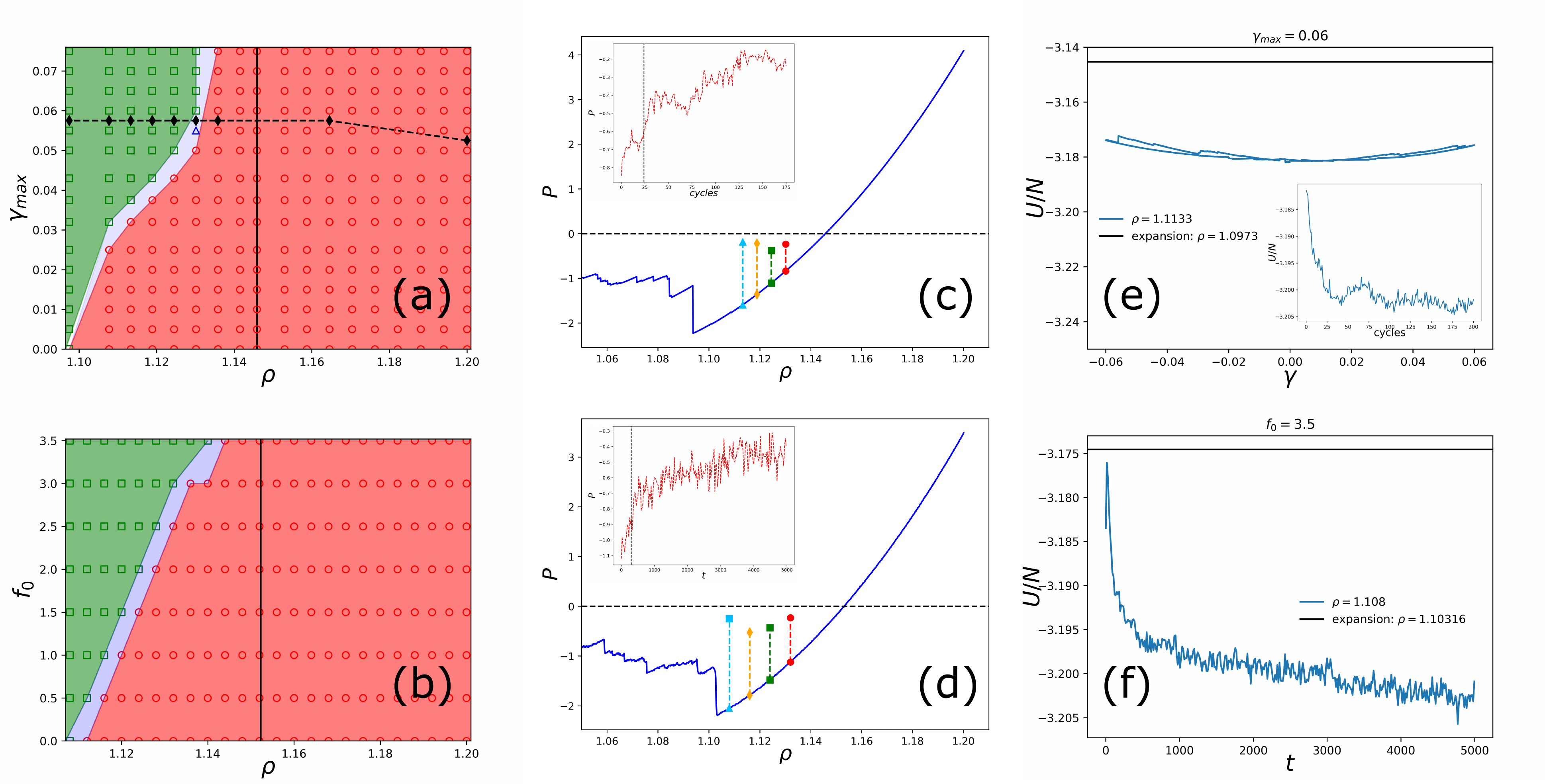}
	\caption{({\bf Left}) Phase diagram corresponding to secondary (a) athermal quasistatic cyclic shear (b) active dynamics at $T=0.01$, imposed on states obtained via expansion imposed on spatially homogeneous amorphous states. The pink sector in the phase diagrams corresponds to spatially homogeneous states and the green sector corresponds to cavitated states. The black vertical line in (a), (b) demarcates the density at which pressure changes sign. Also shown in (a), using diamonds and dashes, is the estimated locus for the onset of irreversible plasticity under applied cyclic shear. ({\bf Middle}) The blue line shows pressure vs density for pure expansion using (c)  athermal quasistatic protocol, and (d) at $T=0.01$. The dotted lines originating from this line and joining circles, squares, diamonds and triangles in both (c) and (d) shows examples of pressure jumps due to cavitation under secondary cyclic shear and activity, respectively, starting from spatially homogenous configurations. The insets in both (c) and (d) show pressure jump, at cavitation, under cyclic shear and active dynamics respectively, for example trajectories, for densities of 1.1133 and 1.108, respectively. ({\bf Right})  For a trajectories marked in (c) and (d), evolution of energy under secondary deformation: (e) for the first cycle under cyclic shear using $\gamma_{max}=0.06$, with inset showing the evolution with cycles, and (f) with time under active dynamics at $T=0.01$ using $f_0=3.5$.}
	\label{fig1}
\end{figure*}

\noindent{\bf Phase diagram: }
Our main result, as discussed above, is that cavitation can be induced earlier, i.e. at higher densities, when the amorphous state is subjected to a secondary drive, be the cyclic shear or the random active forcing. This is summarised via the phase diagrams shown in Fig.\ref{fig1}(a)-(b), where we demonstrate the conditions under which cavitation occurs  in the space of density ($\rho$) and the relevant forcing variable, viz. amplitude of oscillation ($\gamma_{max}$) or  magnitude of active forcing ${f_0}$. Note that in the  phase diagram, the pink sector corresponds to spatially homogeneous states and the green sector corresponding to the cavitated states. It is evident that cavitation occurs at earlier and earlier densities with increasing magnitude of the secondary forcing, for both the cases studied. However, it is important to underline that for cavitation to occur, the ambient pressure has to be negative, i.e. the system has to be in a state of internal tension, which is generated via the expansion.

\vskip +0.1in
\noindent{\bf Energetics: }
Similar to the case of failure via pure expansion, the formation of a cavity via the secondary forcing is also accompanied by a sudden release of the internal pressure; see insets of Fig.\ref{fig1}(c)-(d) for example trajectories in the case of cyclic shear and active dynamics, respectively. In the main panels of Fig.\ref{fig1}(c)-(d), we illustrate at a few densities, via dashed lines, the difference between the initial pressure, prior to the introduction of the second driving, and the eventual pressure attained after the cavitation. It is evident that due to the secondary drive, the cavitation process is occurring at a higher pressure (or lower energy; see Fig.S1 in supplementary information) compared to the pressures (or energies) at which it would occur under pure expansion. In Fig.\ref{fig1}(e), for an example trajectory, we show how the energy evolves during the first cycle after the imposition of cyclic shear and the inset shows how the stroboscopic energy evolves over several cycles; see supplementary information for the data of the full observation window of 200 cycles. Note that, even at maximum strain during the cycle, where maximal plasticity is expected to occur, the system's energy is much less than the threshold needed for cavitation to occur under pure expansion (marked with black line in Fig.\ref{fig1}(e)).  Similarly, in the case of the actively driven system, we observe in Fig.\ref{fig1}(f) that even though energy injection is happening at microscopic scale via the active forcing, the macroscopic potential energy of the system is decreasing, i.e. an annealing process is going on after cavitation and this energy scale is much lower than the situation where cavities appear under pure expansion. These results suggest that secondary deformation
protocols can provide other lower energy (free energy) barrier pathways for relaxation eventually leading to cavity formation in these materials at density which
is not accessible purely via one protocol (i.e. expansion, in this case). In Ref.\cite{krishnan2023annealing}, it has been shown that cyclic shear along multiple orthogonal directions can lead to better annealing than cyclic shear along a single direction. Here, we are able to show that if various different types of deformation, i.e. volumetric expansion and shear (both at bulk and local level), can lead to better annealing of amorphous solids. This also leads to the possibility of achieving stable (ultra stable) glasses using combination of cyclic volumetric and shear deformation in these materials, which is a work in progress. 

\begin{figure}
	\includegraphics[width=0.95\linewidth]{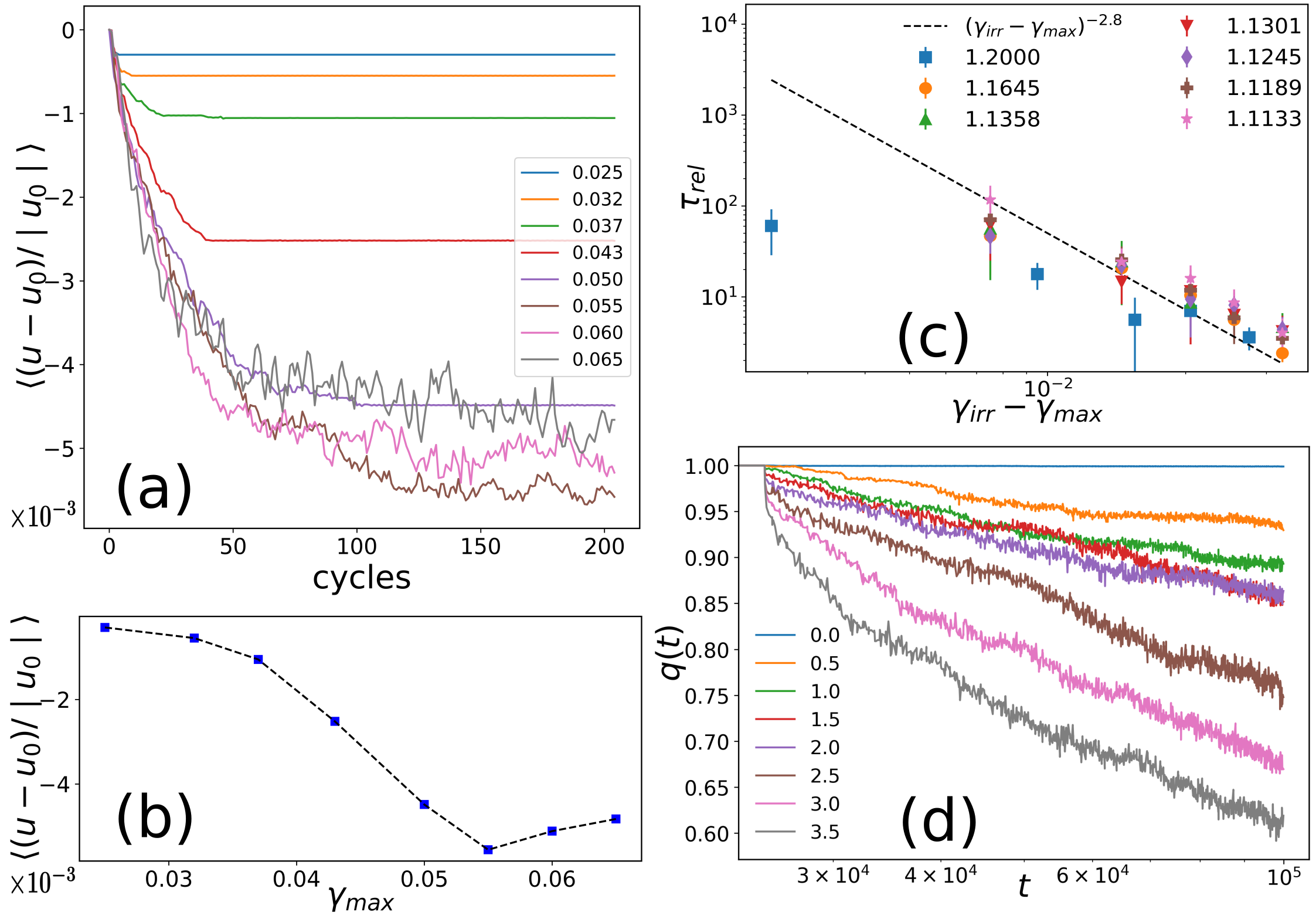}
	\caption{{\bf Secondary cyclic shear: } (a) At a fixed density ($\rho=1.1133$), evolution of rescaled stroboscopic  energies, ${|u-u_0|/|u_0|}$, with number of cycles, for different $\gamma_{max}$ values as labelled.  (b) Corresponding variation of ensemble averaged asymptotic rescaled energies (obtained via averaging across the last 20 cycles)  with shear amplitude. (c) Variation of timescales of relaxation to a limit cycle with $\gamma_{irr} - \gamma_{max}$, for various densities, where $\gamma_{irr}$ is the estimated yielding threshold for large scale plasticity for each density. {\bf Secondary active dynamics: } (d) At density of $\rho=1.108$, variation of overlap function, $q(t)$, with time after the formation of cavity for different values of forcing $f_0$ as labelled.}
	\label{fig3}
\end{figure}

\vskip +0.1in
\noindent{\bf Yielding: }
A pertinent question to ask is whether the cavitation induced by the secondary mechanical perturbation results in large scale plasticity and subsequent 
fluidization. We examine this for both the protocols that we have studied. As already evidenced in Fig.\ref{fig1}(e),(f) for example trajectories, the energy 
decreases after cavitation, both for cyclic shear and active forcing, implying that the system is annealing after the cavity has appeared. We need to check 
across all accessed state points whether the system reaches an arrested or aging glassy state, depending upon athermal or thermal conditions, or whether 
the system has fluid-like behaviour.

In the case of cyclic shear, we probe this by monitoring, across many cycles of the applied shear, the stroboscopically measured normalized energy defined as ${|u-u_0|/|u_0|}$, 
where $u_0$ is the energy at the start of first cycle. A typical scenario, at an example density, is shown in Fig.\ref{fig3}(a). When 
the cyclic shear is imposed, the (normalized) energy of the system starts to decrease with cycles. For the right combinations of densities and amplitudes 
as summarised in the phase diagram, this decrease of energy is accompanied by formation of a cavity. As the solid undergoes cyclic shear, the energy of 
solid decreases in the first few cycles. The decrease could happen due to lack of initial annealing, as previously reported in studies involving high density
glassy systems\cite{SrikanthAnnealing}. Or, it can be due to the formation of a cavity. 
We observe that for smaller $\gamma_{max}$, the system eventually reaches a limit cycle \cite{LimitCycleSrikanth,MaloneyYielding,DheerajKumar,krishnan2023annealing} 
where either the energy decreases to a constant value or oscillates between two or more different values. Examples of limit cycles with a period greater than one are shown in the Supplementary Information. For larger amplitudes of the oscillatory shear,  the system becomes diffusive over observation timescales. Beyond a certain amplitude, the average energy of the relaxed solid averaged over last few cycles 
starts to increase; see Fig.\ref{fig3}(b). The cusp marks the yielding point under cyclic shear\cite{SrikanthPRX,SrikanthAnnealing}. The yield threshold of the lower density solid 
is higher than the yield threshold of the homogeneous solid at the beginning of expansion; the corresponding locus of yield thresholds is marked in the 
phase diagram of Fig.\ref{fig1}(a) (using diamonds connected with dashes).  
Further, we track the time taken for a system to reach a limit cycle $\tau_{rel}$, as a function of amplitudes is shown in Fig.\ref{fig3}(c) at different densities. 
The power law observed in the works of Kumar et al\cite{DheerajKumar}, $\tau_{rel}\sim(\gamma_{irr}-\gamma)^{-2.8}$ here too serves as a good guide 
to the eye; $\gamma_{irr}$ is the estimated yield threshold under cyclic shear.

For the case of imposed active dynamics, we do a similar check for large scale plasticity by computing the overlap function, $q(t)$, relative to a 
post-cavitated state. If the system is fluidized, then the overlap function should completely decorrelate within observation time. However, as observed in 
Fig.\ref{fig3}(d), this is not the case, implying that the system, even in the presence of activity, is still in the glassy state, for the range of active forcing that 
we have explored. It is possible that higher activity strengths will lead to fluidization, as has been reported earlier for dense glassy 
systems\cite{mandal2016}. 

\begin{figure*}
	\includegraphics[width=0.98\linewidth]{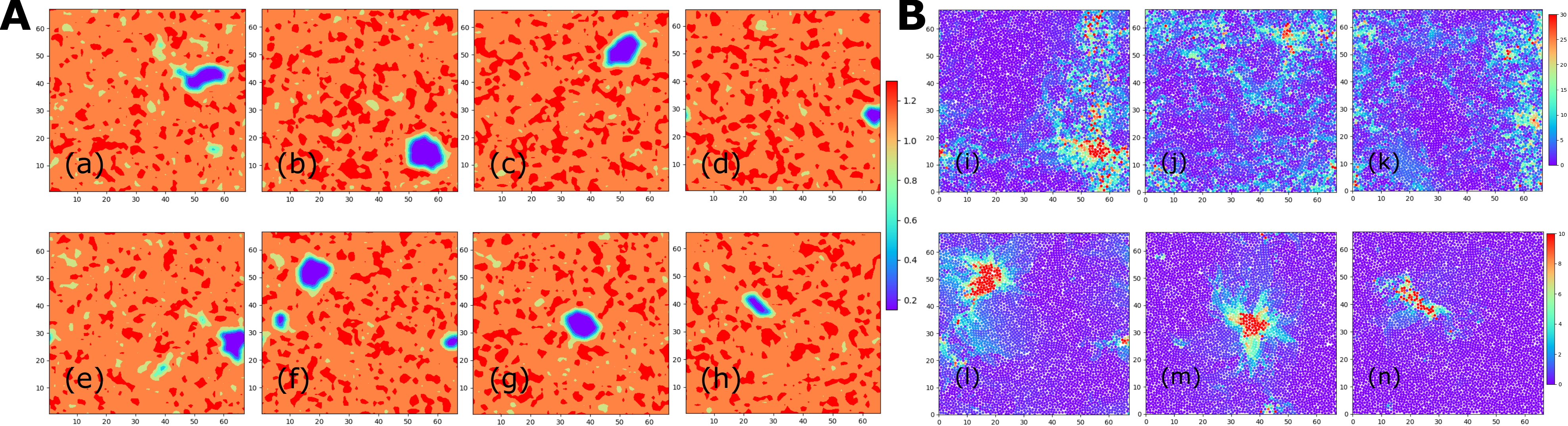}
	\caption{({\bf A}) Maps of coarse-grained density field.  (a) and (e): cavitated states due to pure expansion, occurring at $\rho=1.097$ and $1.1028$, respectively, under athermal quasistatic conditions and at finite temperature ($T=0.01$). Along corresponding trajectories, density maps of cavitated steady state under secondary drive:  (b)-(d) cyclic shear using $\gamma_{max}=0.06$ and shown for $\rho=1.119, 1.124, 1.130$, (f)-(h) active dynamics using $f_0=3.5$ and shown for $\rho=1.108, 1.116, 1.132$.  ({\bf B}) Corresponding spatial maps of squared displacements relative to initial state (i.e. prior to application of secondary deformation), for (i)-(k)  cyclic shear and (l)-(n) active dynamics.}
	\label{fig2}
\end{figure*}

\vskip +0.1in
\noindent{\bf Spatial manifestations: }
Finally, we examine the spatial structures emerging via cavitation induced by the secondary deformation protocol.  We construct maps of the 
coarse-grained density field  \cite{TestardBerthierKob}. In Fig.\ref{fig2}(A), we show these density maps, after cavitation has occurred, corresponding 
to the cyclic shear (top panel) and the active forcing (bottom panel), for some of the density points marked in Fig.\ref{fig1}(c)-(d).  We find that, although 
the initial states in each case are sampled from the same expansion trajectory,  these cavities occur at different locations in the solid, for both cyclic shear 
and active dynamics, if we compare across densities. This suggests abundance of potential cavitation sites present throughout the solid and also multiple 
possible pathways on the energy landscape for the cavitation to occur.   In supplementary material (see Fig.S2 and Fig.S3), we provide illustrative panels 
spanning across density and forcing, which show how the density maps reveal occurrence  of cavities, at different locations, depending upon the state 
point, once the necessary forcing threshold has been crossed at each density, providing a pictorial snapshot of the phase diagrams shown in Fig.\ref{fig1}. 

Further, to characterize the extent of deformation undergone between the initial state, prior to the application of the secondary drive, and the final state 
achieved during the observation time window after the application of the drive, we construct spatial maps of mobility (See Methods). 
In Fig.\ref{fig2}(B), we show these maps for the cases in Fig.\ref{fig2}(A) where cavitation is observed at finite secondary drive, viz. subplots (b)-(d) for 
cyclic shear and (f)-(h) for active dynamics. We observe large displacements around the location where cavity is formed for both the secondary 
deformation protocols. In cyclic shear, for large $\gamma_{max}$, we see extended spatial structures formed by the regions undergoing large displacements,
as shown in Fig.\ref{fig2}(B) (i)-(k), whereas for smaller $\gamma_{max}$, these structure are more localised around the cavity; see Supplementary Information. Thus, as $\gamma_{irr}$ is approached, the spatial scale of plasticity gets more expanded and one expects these to take avalanche-like structures around yield. In the case of active dynamics, where we continue to remain glassy and not fluidized within the scale of forcings explored, the particles with large displacements seem to be concentrated in localised clusters around the cavity; see Fig.\ref{fig2}(B) (l)-(n) (and Supplementary Information). Perhaps, here too, we can expect extended structures at larger forcings.




\section{Discussion}


In summary, using numerical simulations, we have provided proof-of-concept demonstration that in comparison to a case of pure axial expansion, cavitation can occur at smaller densities when the amorphous solid, expanded from a  high density state, is subjected to secondary mechanical deformations. We have shown this phenomenon for two examples of secondary deformation, viz. cyclic shear and active dynamics. Our preliminary analysis (not shown) for uniform shear also reveals similar behaviour. Thus, it shows how combination and interplay of different deformation modes can lead to early onset of cavitation, which evidently would lead to earlier fracture if the expansion process continues simultaneously. However, note that, for cavitation to happen, the overall pressure of the system has to be negative, even when secondary forcing is at play. Thus, the presence of negative tensile stress that emerged from the strain caused via the expansion is necessary for the secondary shearing to couple with in order for the relaxation to happen in the form of cavity formation. Overall, our work  has significant implications for real-life scenarios where typical applications comprise of complex loading mechanisms which is a mixture of multiple deformation modes, and  thus materials have to be appropriately designed to prevent failure via such cross-coupling.  Further, our study paves the way for more explorations of probing failure in amorphous solids via complex deformations, across the entire spectrum of hardness that these materials span. 


One of the interesting findings of the work is that the cavitated state that one obtains via the secondary deformation could be an arrested or glassy state, depending upon whether the system is athermal or thermal. Or, in other words, the secondary deformation is providing a pathway to access these local minima in the energy landscape, at densities where usual axial expansion would not find them. Further, we demonstrate that multiple local
minima, manifested in the form of different structurally inhomogeneous states, emerge along the same trajectory, depending upon the density
and the strength of the secondary forcing. This implies that different soft spots get selected for the cavitation to occur, via the coupling
between the different deformation modes and their relative strengths.

In the context of active dynamics, the emergence of cavities reminds
of motility induced phase separation (MIPS) reported in the active matter systems \cite{cates2015motility}, albeit the minority phase being the voids rather than the more well-studied active droplet. But this would be expected in the high density limit \cite{villarroel2021critical, PhysRevE.106.L012601}. In most studies related to MIPS, the active particles interact via repulsive interaction and the effective attraction emerges out of the active dynamics. In our case, all the particles have Lennard-Jones interactions, including the small fraction of particles that are active.
The picture that we have in mind is that these small number of active particles act as random local sources of shear which collectively organise to help in relaxing a mode that eventually leads to cavitation. Thus, it is also possible that the MIPS formation in the usual active systems also has a shearing mechanism at its origin. 

In recent times, significant progress has been made towards the development of simple mesoscale models that use coarse-grained variables like stress and strain to provide insight into the physical processes at play during the response exhibited by amorphous solids to mechanical loadings. Not only have these models been successful in reproducing the phenomenology \cite{nicolas2018deformation} but also have been able to bridge towards microscopic models \cite{liu2021elastoplastic}. However, most of the modelling has so far focussed on the response to applied shear. Our work on the coupling of various deformation modes and its consequence to failure via cavitation should now motivate further development of these models towards more complex loading scenarios.

\section{Acknowledgements}

{We thank the HPC facility at IMSc-Chennai and at TIFR-H for providing computational resources. S.K. would like to acknowledge the support from Swarna Jayanti Fellowship Grants No. DST/SJF/PSA-01/2018-19 and No.  SB/SFJ/2019-20/05 and Core Research Grant from SERB via grant CRG/2019/005373. PC acknowledges funding from SERB via grant MTR/2022/001034. We also thank Peter Sollich for useful discussions.}

\bibliography{references}

\end{document}